\documentclass[prb,fleqn,12pt,showpacs]{revtex4}
\usepackage{epsfig}
\usepackage{graphicx}
\usepackage{amsfonts}
\begin{document}
\title{Spin-charge and spin-orbital coupling effects \\
on spin dynamics in ferromagnetic manganites}
\author{Dheeraj Kumar Singh, Bhaskar Kamble, and Avinash Singh}
\email{avinas@iitk.ac.in}
\affiliation{Department of Physics, Indian Institute of Technology Kanpur}
\begin{abstract}
Correlation-induced spin-charge and spin-orbital coupling effects on spin dynamics in ferromagnetic manganites are calculated with realistic parameters in order to provide a quantitative comparison with experimental results for spin stiffness, magnon dispersion, magnon damping, anomalous zone-boundary magnon softening, and Curie temperature. The role of orbital degeneracy, orbital ordering, and orbital correlations on spin dynamics in different doping regimes is highlighted. 
\end{abstract}
\maketitle

\newpage
\section{Introduction}
The role of charge and orbital fluctuations on magnetic couplings and excitations is of strong current interest in view of the several zone-boundary anomalies observed in spin-wave excitation measurements in the metallic ferromagnetic phase of colossal magnetoresistive manganites.\cite{hwang_98,dai_2000,tapan_2002,ye_2006,ye_2007,zhang_2007,moussa_2007} The presence of short-range dynamical orbital fluctuations has been suggested in neutron scattering studies of ferromagnetic metallic manganite $\rm La_{1-x}(Ca_{1-y}Sr_y)_{x}MnO_3$.\cite{moussa_2007} These observations are of crucial importance for a quantitative understanding of the carrier-induced spin-spin interactions, magnon excitations, and magnon damping, and have highlighted possible limitations of existing theoretical approaches. 

For example, the observed magnon dispersion in the $\Gamma$-X direction shows significant softening near the zone boundary, indicating non-Heisenberg behaviour usually modeled by including a fourth neighbour interaction term $J_4$, and highlighting the limitation of the double-exchange model. Similarly, the prediction of magnon-phonon coupling as the origin of magnon damping\cite{dai_2000} and of disorder as the origin of zone-boundary anomalous softening\cite{furukawa_2005} have been questioned in recent experiments.\cite{ye_2006,ye_2007,zhang_2007,moussa_2007} Furthermore, the dramatic difference in the sensitivity of long-wavelength and zone-boundary magnon modes on the density of mobile charge carriers has emerged as one of the most puzzling feature. Observed for a finite range of carrier concentrations, while the spin stiffness remains almost constant, the anomalous softening and broadening of the zone-boundary modes show substantial enhancement with increasing hole concentration.\cite{ye_2006,ye_2007}

The role of orbital degeneracy, orbital ordering and orbital correlations on spins dynamics in ferromagnetic manganites in the entire  hole-doping range $0 < x \lesssim 0.5$ is still not fully understood due to lack of quantitative comparisons of theoretical calculations with experimental results in manganites, as highlighted in the recent review.\cite{ye_2007}

Most of the earlier investigations including correlation-induced O$(1/S)$ quantum corrections to the magnon spectrum were carried out in the strong-coupling (double-exchange) limit and could not satisfactorily account for the observed anomalous zone-boundary magnon softening.\cite{kapetanakis_2006} Whereas, with typical parameter values $t \sim 0.2-0.5$ eV for the mobile $e_g$ electrons
and $2J \sim 2$ eV for their exchange coupling to the localized $t_{2g}$ core spins, the intermediate-coupling regime $J \sim W$ appears to be more appropriate for the ferromagnetic manganties.\cite{dagotto_2001}

Recent theoretical studies in the intermediate-coupling regime and including the Coulomb repulsion between band electrons have demonstrated some realistic features such as doping dependent asymmetry of the ferromagnetic phase and enhanced magnon softening and non-Heisenberg behaviour, thereby highlighting the importance of correlated motion of electrons on spin dynamics.\cite{golosov_2005,kapetanakis_2007} However, these investigations do not include orbital degeneracy and inter-orbital Coulomb interaction, and do not provide any quantitative comparisons with experimental results on manganites. Earlier theoretical investigations of the role of orbital-lattice fluctuations and correlations on magnetic couplings and excitations have mostly been limited to ferro orbital correlations,\cite{khaliullin_2000} and have not addressed the physically relevant staggered orbital correlations. 

A self-consistent investigation of the interplay between spin and orbital orderings has recently carried out using a two-orbital ferromagnetic Kondo lattice model (FKLM) including the Jahn-Teller coupling,\cite{stier_2007} and doping dependence of the calculated Curie temperature was compared with experiments for different ferromagnetic manganites. Finite Jahn-Teller distortion and orbital ordering were shown to be self-consistently generated at low doping. However, again only ferro orbital ordering was included in this analysis. 

Recent investigations of the correlated motion of electrons using a non-perturbative, inverse-degeneracy-expansion based, Goldstone-mode-preserving approach have highlighted the role of spin-charge and spin-orbital coupling effects on magnetic couplings and excitations in orbitally degenerate metallic ferromagnets.\cite{singh,spch3,qfklm,sporb} 
While spin-charge coupling was indeed found to yield strong magnon softening, damping, and non-Heisenberg behaviour, it was only on including orbital degeneracy, inter-orbital interaction, and a new class of spin-orbital coupling diagrams that 
low-energy staggered orbital fluctuations, particularly with momentum near $(\pi/2,\pi/2,0)$ corresponding to CE-type orbital correlations, were found\cite{sporb} to generically yield strong intrinsically non-Heisenberg $(1-\cos q)^2$ magnon self energy correction in three dimensions, resulting in no spin stiffness reduction, but strongly suppressed zone-boundary magnon energies in the $\Gamma$-X direction. 

In this paper we will apply these new results specifically to the case of ferromagnetic manganites. We will consider a two-orbital FKLM including intra- and inter-orbital Coulomb interactions, with realistic values of hole bandwidth, lattice parameter, and Hund's coupling etc. Such a quantitative comparison should provide, within a physically transparent theoretical approach, insight into the role of orbital ordering, orbital degeneracy, and orbital correlations on spin dynamics in ferromagnetic manganites, seamlessly covering the entire  hole-doping range $0 < x \lesssim 0.5$ within a single theoretical framework. 

The outline of the paper is as follows. Spin-charge coupling effects on spin dynamics in a single-band FKLM are presented in Section II, highlighting the strong differences from the canonical double-exchange behaviour. Spin stiffness and spin-wave energies are then obtained in the orbitally degenerate state of a two-orbital FKLM in Section III, and compared with neutron-scattering results for the optimally ferromagnetic manganites $(x \approx 0.3)$. Inter-orbital interaction $V$ is included next and Section IV describes spin dynamics in the orbitally ordered ferromagnetic state. The hole-doping $(x)$ dependence of the estimated Curie temperature is compared with experimental results for two different cases corresponding to wide- and intermediate-band manganites. Finally, the role of $(\pi/2,\pi/2,0)$-type orbital correlations on zone-boundary magnon softening and instability of the FM state near $x=0.5$ are discussed in section V. 

\section{Spin-charge coupling magnon self energy} 

The interplay between itinerant carriers in a partially filled band and localized magnetic moments is conventionally studied within the ferromagnetic Kondo lattice model (FKLM):
\begin{equation}
H = -t \sum_{\langle ij\rangle \sigma} a^\dagger _{i\sigma}  a_{j\sigma}
- J \sum_i {\bf S}_{i}.{\mbox{\boldmath $\sigma$}}_i 
\end{equation}
involving a local exchange interaction between the localized spins ${\bf S}_i$ and itinerant electron spins ${\mbox{\boldmath $\sigma$}}_i$. Due to crystal-field splitting of the degenerate Mn $3d$ orbitals into $e_g$ and $t_{2g}$ levels, the partially filled band in ferromagnetic manganites corresponds to the mobile $e_g$ electrons whereas the three Hund's-coupled localized $t_{2g}$ electrons yield the localized magnetic moments with spin quantum number $S=3/2$.  

We will first consider the single-band FKLM, which is appropriate for the orbitally degenerate state of the optimally doped ferromagnetic manganites near hole doping $x\approx 0.3$. As the doped holes are shared equally by the two degenerate orbitals, this hole doping corresponds to band filling $n = (1-x)/2 \approx 0.35$ in each band. Orbital ordering and orbital correlations will be incorporated later  within a two-orbital FKLM including intra-orbital and inter-orbital Coulomb interactions. Throughout, we will consider a simple cubic lattice with lattice parameter $a$.  

\begin{figure}
\vspace*{-20mm}
\hspace*{-0mm}
\psfig{figure=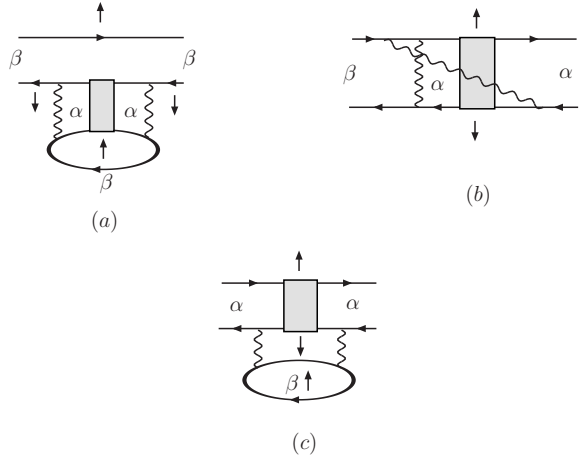,width=120mm}
\vspace{-100mm}
\caption{The first-order quantum corrections to the irreducible particle-hole propagator $\phi({\bf q},\omega)$.}
\end{figure}

The first-order magnon self energy for the single-band FKLM, resulting from quantum corrections to the irreducible particle-hole propagator $\phi({\bf q},\omega)$ shown in Fig. 1, was recently obtained as:\cite{qfklm}
\begin{eqnarray}
\Sigma_{\rm magnon} ({\bf q},\omega) &=& J^2 (2S) \sum_{\bf Q} \int \frac{d\Omega}{2\pi i} \; 
\left ( \frac{1}{\Omega + \omega_{\bf Q}^0 -i \eta} \right ) \sum_{\bf k} \left [ \left ( \frac{1}
{\epsilon_{\bf k-q+Q}^{\uparrow +} -\epsilon_{\bf k}^{\uparrow -} + \omega -\Omega -i \eta} \right ) \right . \nonumber \\
& \times & \left . \frac{1}{2S} \left ( 1-\frac{2JS}
{\epsilon_{\bf k-q}^{\downarrow +} -\epsilon_{\bf k}^{\uparrow -} + \omega -i \eta} \right )^2 \right ]
\end{eqnarray}
where $\omega_{\bf Q}^0$ refers to the bare magnon energies, $\epsilon_{\bf k}^\sigma = \epsilon_{\bf k} - \sigma JS$ are the exchange-split band energies, and the superscripts $+/-$ indicate particle ($\epsilon_{\bf k}^\sigma > \epsilon_{\rm F}$) and hole ($\epsilon_{\bf k}^\sigma < \epsilon_{\rm F}$) energies. Incorporating correlation-induced self-energy and vertex corrections, the above magnon self energy represents a spin-charge coupling between magnons and charge excitations in the partially-filled majority-spin band, with the coupling vertex explicitly vanishing for $q=0$, thus ensuring that the Goldstone mode is explicitly preserved. This result was obtained by applying the systematic inverse-degeneracy $(1/{\cal N})$ expansion scheme to a purely fermionic representation for the FKLM which allows conventional many-body diagrammatic expansion. 

The imaginary part of the magnon self energy in the strong-coupling limit:
\begin{equation}
\frac{1}{\pi} {\rm Im} \Sigma_{\rm magnon} ({\bf q},\omega) = \left ( \frac{1}{2S}\right )^2 
\sum_{\bf Q} \sum_{\bf k} (\epsilon_{\bf k-q} - \epsilon_{\bf k} + \omega)^2  
\delta(\epsilon_{\bf k-q+Q}^{\uparrow +} -\epsilon_{\bf k}^{\uparrow -} +\omega +\omega_{\bf Q}^0 ) 
\end{equation}
yields finite magnon damping and linewidth at zero temperature, arising from magnon decay into intermediate magnon states accompanied with majority-spin charge excitations. On the other hand, at the classical level (random phase approximation), magnon damping is possible only due to decay into the Stoner continuum, and is therefore completely absent in the intermediate and strong coupling regimes where the magnon spectrum lies well within the Stoner gap.

\begin{figure}
\begin{center}
\vspace*{-2mm}
\hspace*{0mm}
\psfig{figure=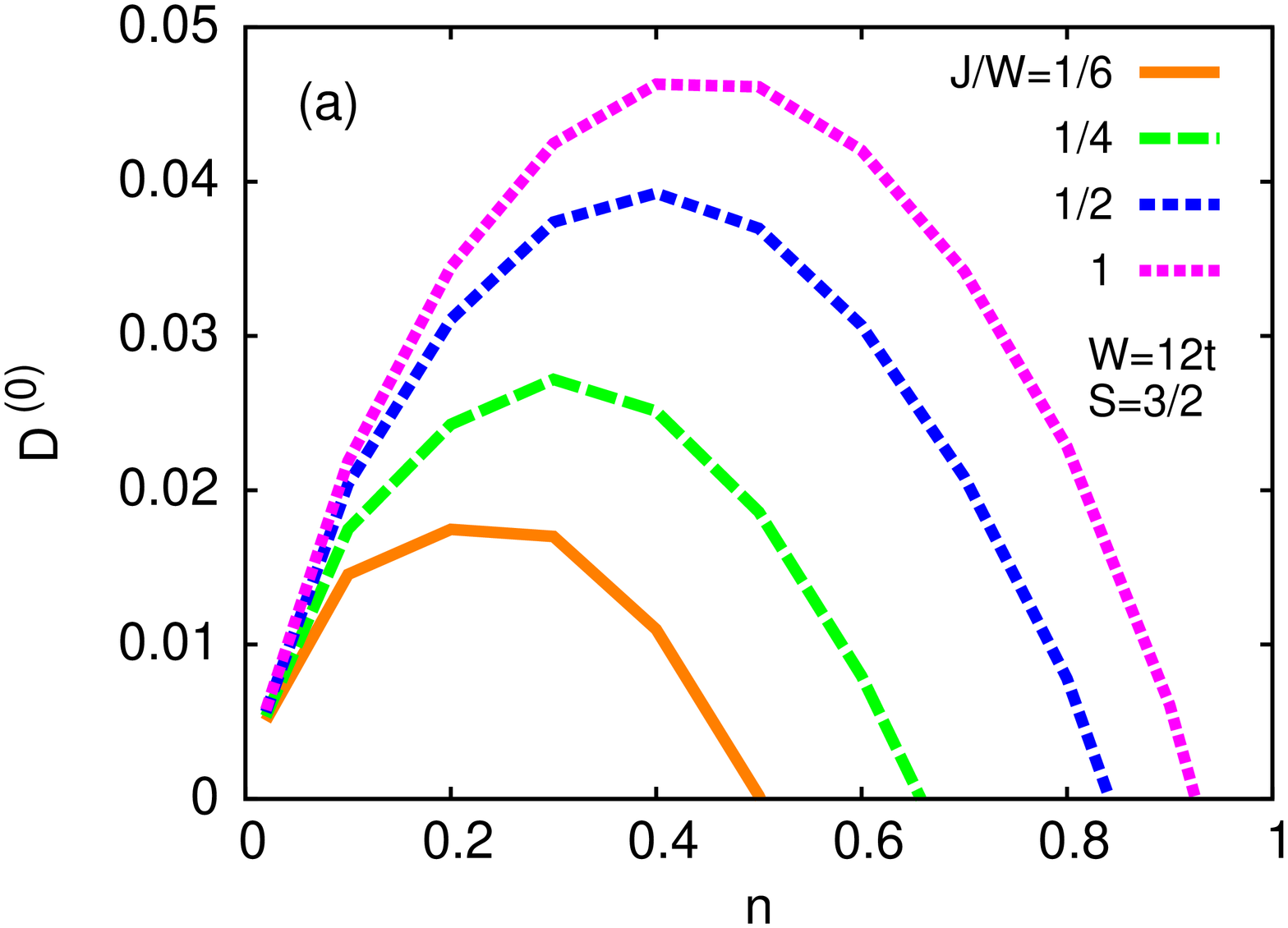,width=55mm,angle=-90}
\psfig{figure=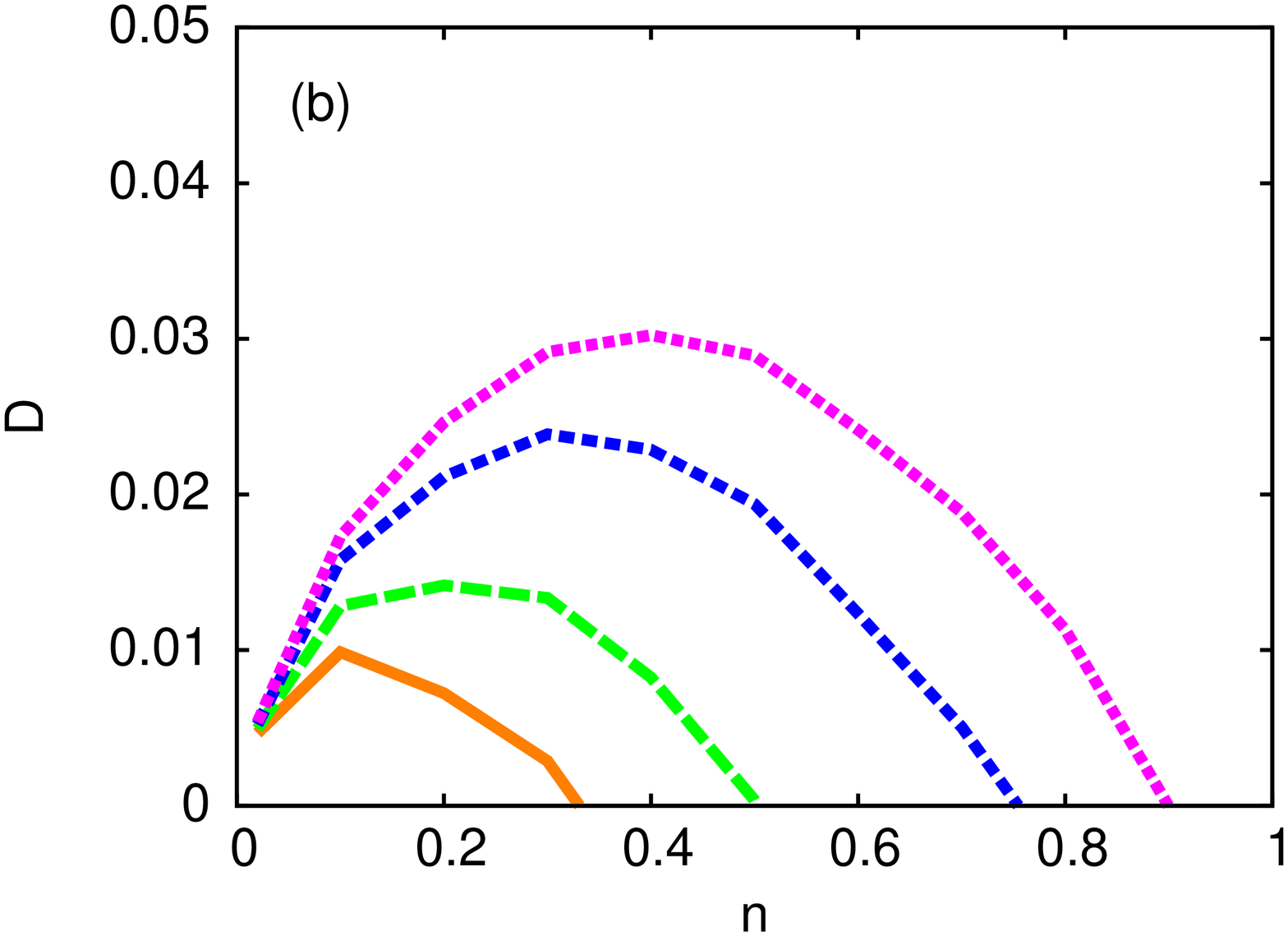,width=55mm,angle=-90}
\vspace*{-5mm}
\end{center}
\caption{Comparison of the bare (a) and renormalized (b) spin stiffness for the single-band FKLM, showing the substantial spin stiffness reduction due to the spin-charge coupling effect.}
\label{fig1_dks}
\end{figure}

The small-$q$ behaviour of the magnon self energy in Eq. (2) yields the first-order quantum correction to spin-stiffness in $d$ dimensions:
\begin{equation}
D^{(1)} = \Sigma^{(1)} ({\bf q})/q^2 = \frac{1}{d(2S)^2}  \sum_{\bf Q,k} 
\frac{({\mbox{\boldmath $\nabla$}}\epsilon_{\bf k})^2}
{\epsilon_{\bf k-q+Q}^{\uparrow +} -\epsilon_{\bf k}^{\uparrow -}+\omega_{\bf Q} ^0} 
\label{ren_stiff}
\end{equation}
which is down by a factor $(1/2S)$ relative to the bare (classical) spin stiffness: 
\begin{equation}
D^{(0)} = \omega_{\bf q} ^{(0)} / q^2 = \frac{1}{d(2S)} \left [\frac{1}{2} 
\langle {\mbox{\boldmath $\nabla$}}^2 \epsilon_{\bf k} \rangle  - 
\frac{\langle ({\mbox{\boldmath $\nabla$}} \epsilon_{\bf k})^2 \rangle }{2JS} \right ] \; .
\end{equation}
The two competing terms above of order $t$ and $t^2/J$ correspond to delocalization energy loss and exchange energy gain upon spin twisting, and determine the overall stability of the ferromagnetic state against long wavelength fluctuations. 
The spin stiffness quantum correction $D^{(1)}$ is only weakly dependent on $J$ (through $\omega_{\bf Q} ^0$), and involves only an exchange contribution, which is suppressed for electronic spectral distributions characterized by dominant saddle-point behaviour (${\mbox{\boldmath $\nabla$}}\epsilon_{\bf k} = 0$), resulting in enhanced spin stiffness 
$D = D^{(0)} - D^{(1)}$ and ferromagnetic stability. 

In the double-exchange limit ($J\rightarrow \infty$), only the delocalization part of the classical spin stiffness survives. This contribution has a particle-hole symmetric parabolic band-filling dependence. Its behaviour with electron filling $n$ is identical to that with hole doping $x$, vanishing at both ends $n=0$ and $n=1$ and increasing symmetrically with added electrons or holes, leading to the conventional understanding within the DE model that ferromagnetism increases with added carriers. However, both the classical and quantum exchange contributions involving the 
$({\mbox{\boldmath $\nabla$}}\epsilon_{\bf k})^2$ terms in Eqs. (4) and (5) break this particle-hole symmetry, and the stiffness actually vanishes at $n < 1$ before the band is completely filled.    


\begin{figure}
\begin{center}
\vspace*{-2mm}
\hspace*{0mm}
\psfig{figure=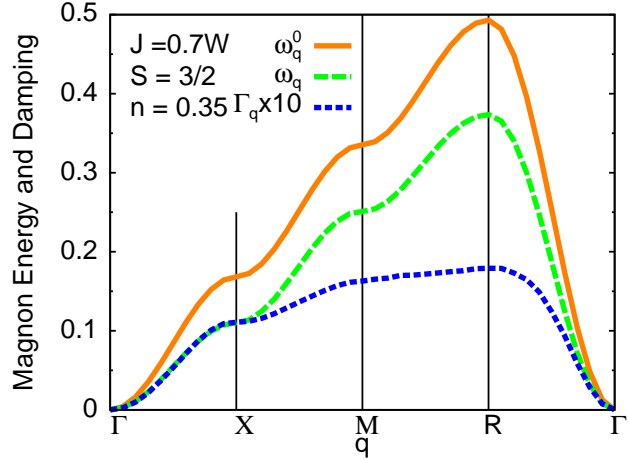,width=60mm,angle=-90}
\vspace*{-5mm}
\end{center}
\caption{The bare and renormalized magnon energies and magnon damping for the single-band FKLM. The magnon damping linewidths are nearly one-tenth of the renormalized magnon energies in the $\Gamma$-X direction.}
\label{fig2_dks}
\end{figure}

Since the spin stiffness quantum correction involves only an exchange contribution, including quantum correction is therefore effectively equivalent to enhancing the bare exchange contribution by decreasing $J$. Fig. \ref{fig1_dks} shows the renormalized spin stiffness $D = D^{(0)} - D^{(1)}$ in units of $ta^2$, evaluated from Eqs. (4) and (5) for different values of $J$. For band fillings below $n=0.5$, the renormalized spin stiffness for $J \sim W$ is indeed seen to be close to the bare spin stiffness for $J/t=4$. 

Figure \ref{fig2_dks} shows a comparison of the bare and renormalized magnon dispersions in units of $t$, and also the magnon damping obtained for the single-band FKLM. Besides the substantial reduction of the renormalized magnon energies, we also find that  in the intermediate coupling regime ($J \sim W$), the magnon self energy becomes increasingly non-Heisenberg like, resulting in $\omega_X < \omega_R/3$, as indeed observed. The magnon linewidth $\Gamma_{\bf q}$ obtained from the imaginary part of the magnon self energy increases with momentum $q$, but the ratio $\Gamma_{\bf q}/\omega_{\bf q}$ is found to remain nearly constant at $\sim 1/10$ upto the zone boundary in the $\Gamma$-X direction, in agreement with magnon linewidth measurements.\cite{moussa_2007} In the X-M, M-R, and R-$\Gamma$ directions, we find the ratio $\Gamma_{\bf q}/\omega_{\bf q}$ to be smaller than 0.1.

\begin{figure}[htp]
\begin{center}
\vspace*{-2mm}
\hspace*{0mm}
\psfig{figure=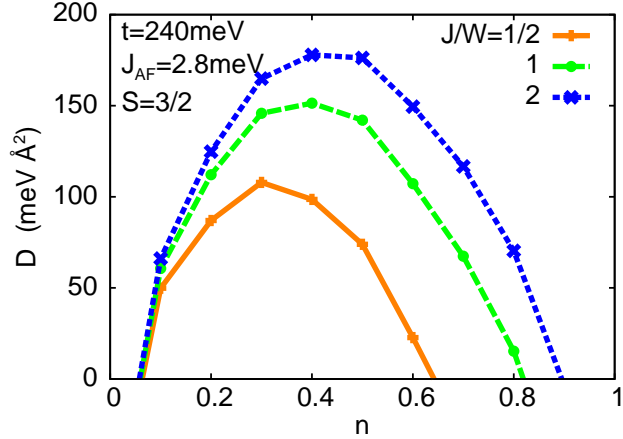,width=60mm,angle=-90}
\vspace*{-5mm}
\end{center}
\caption{The renormalized spin stiffness calculated for a two-orbital model, including an AF contribution due to the Mn superexchange. For $J \approx W$, the calculated values are close to the measured stiffness $\sim 170$ meV\AA$^2$ for the wide-band compound $\rm La_{1-x} Sr_{x} Mn O_3$.}
\label{fig3_dks}
\end{figure}

The spin-charge coupling effect on magnon excitations can be readily extended to the two-band FKLM involving interaction term 
$- J \sum_i {\bf S}_{i}.({\mbox{\boldmath $\sigma$}}_{i\alpha} + {\mbox{\boldmath $\sigma$}}_{i\beta})$ in Eq. (1). 
Fig. \ref{fig3_dks} shows the renormalized spin stiffness for this two-band model obtained by doubling the spin stiffness result $D = D^{(0)} - D^{(1)}$ from Eqs. (4,5), corresponding to the independent contributions of the two degenerate $e_g$ orbitals. Here we have taken $t=240$meV and lattice parameter $a=3.87$\AA\ for manganites. Corresponding to the AF superexchange interaction between Mn spins, an AF contribution to spin stiffness $D^{(0)}_{\rm AF} = zJ_{\rm AF} S a^2/6$ was also subtracted from the above result with $J_{\rm AF}=2.8$meV and lattice coordination $z=6$. 

Near optimal band filling $n \approx 0.35$ (hole doping $x = 1-2n \approx 0.3$) for ferromagnetic manganites, the calculated spin stiffness for $J \approx W$ is in close agreement with the experimentally measured values $ \sim 170$ meV\AA$^2$ for the wide-band compound $\rm La_{1-x} Sr_{x} Mn O_3$. The strong suppression near n=0.5 (x=0) seen above with decreasing $J$ could also be a contributing factor towards destabilizing ferromagnetism in manganites as $x$ approaches zero, in addition to orbital correlations.

\begin{figure}
\begin{center}
\vspace*{-2mm}
\hspace*{0mm}
\psfig{figure=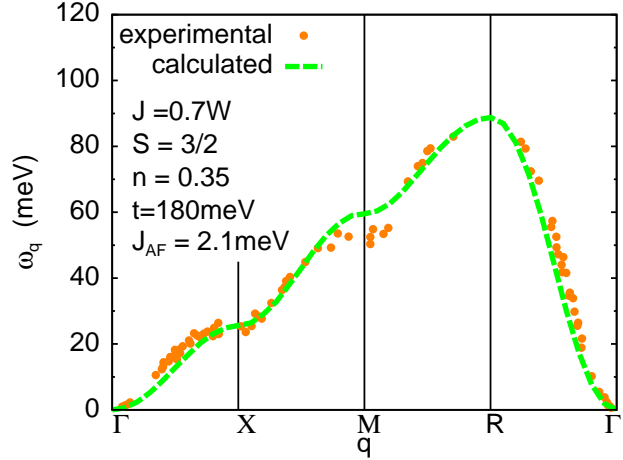,width=60mm,angle=-90}
\vspace*{-5mm}
\end{center}
\caption{Comparison of calculated renormalized magnon energy dispersion with experimental data for the narrow-band compound $\rm La_{0.7} Ca_{0.3} Mn O_3$ from neutron scattering measurements.\cite{ye_2007}}
\label{fig4_dks}
\end{figure}

Fig. \ref{fig4_dks} shows a comparison of the calculated magnon dispersion for the two-band FKLM along different symmetry directions in the Brillouin zone with experimental data for the narrow-band compound $\rm La_{0.7} Ca_{0.3} Mn O_3$ obtained from neutron scattering measurements.\cite{ye_2007} The renormalized magnon energy was obtained using 
$\omega_{\bf q} = 2(\omega_{\bf q}^0 - \Sigma_{\bf q}) - zJ_{\rm AF} S (1-\gamma_{\bf q})$. A relatively smaller hopping term $t=180$meV (bandwidth $W=12t\approx 2$eV) was taken in the calculation for this narrow-band compound. Ab-initio calculations also yield an estimated bandwidth $W=2$eV for $\rm La_{1-x} Ca_x MnO_3$.\cite{trimarchi_2005} The renormalized magnon energies are broadly in good agreement with the neutron scattering measurements. Including the zone-boundary magnon softening resulting from the spin-orbital coupling effect discussed later will improve the agreement near the X point. 

\begin{figure}[htp]
\begin{center}
\vspace*{-2mm}
\hspace*{0mm}
\psfig{figure=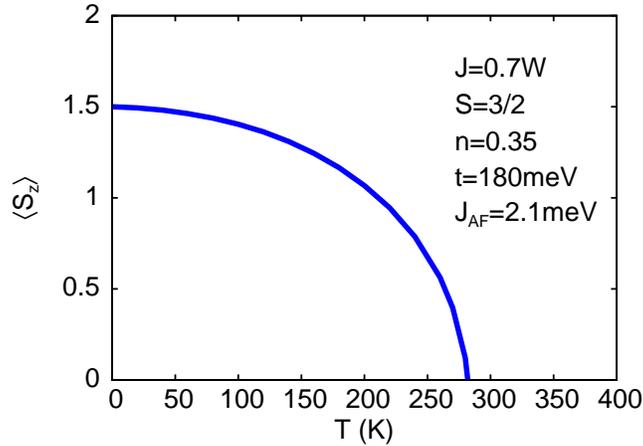,width=60mm,angle=-90}
\vspace*{-5mm}
\end{center}
\caption{Temperature dependence of magnetization calculated using the self-consistent Callen scheme with the same set of parameters as above. The calculated Curie temperature is close to the measured values for narrow-band manganites.}
\label{fig5_dks}
\end{figure}

From the renormalized magnon energies obtained as above, the finite-temperature spin dynamics was investigated using the self-consistent Callen scheme in which the magnetization for a quantum spin-$S$ ferromagnet:\cite{callen}
\begin{equation}
\langle S^z\rangle=\frac{(S-\Phi)(1+\Phi)^{2S+1}+(S+1+\Phi)\Phi^{2S+1}}{(1+\Phi)^{2S+1}-\Phi^{2S+1}} \; ,
\end{equation}
where the boson (magnon) occupation number:
\begin{equation}
\Phi = \frac{1}{N} \sum_{\bf q} \frac{1}{e^{\beta\tilde{\omega}_{\bf q}} - 1}
\end{equation}
in terms of the thermally renormalized magnon energies:
\begin{equation}
\tilde{\omega}_{\bf q} = \omega_{\bf q} \langle S_z\rangle/S 
\end{equation}

Figure \ref{fig5_dks} shows the temperature dependence of magnetization obtained by self consistently solving the coupled set of equations (6)-(8). With the same set of parameters as in Fig. 5 for the magnon dispersion fit, the calculated Curie temperature ($\sim \rm 280K$) is close to the measured value $\sim \rm 250K$ for LCMO.


As orbital correlations are relatively unimportant at these fillings, the orbitally degenerate FKLM calculation presented here provides a good description of ferromagnetic properties. Orbital ordering and fluctuations due to inter-orbital interaction and Jahn-Teller distortion become important near quarter filling ($x \sim 0$) and near $x \sim 0.5$, and therefore must be included as both strongly suppress ferromagnetism. Effects of orbital ordering and orbital fluctuations on spin dynamics are discussed below.

\section{orbitally ordered ferromagnetic state}

Weakly doped manganites such as $\rm La_{1-x} Sr_x Mn O_3$ and $\rm La_{1-x} Ca_x Mn O_3$ exhibit an orbitally ordered state for $x \lesssim 0.2$, as inferred from x-ray diffraction and neutron scattering experiments.\cite{pissas_2005,moussa_2007} 
In this section, we will therefore consider spin dynamics in an orbitally ordered ferromagnet with staggered orbital ordering, and examine the behaviour of the Curie temperature with the onset of orbital ordering in the low-doping regime. We therefore consider a two-orbital correlated FKLM:

\begin{equation}
H = -t \sum_{\langle ij \rangle \sigma\mu} a^\dagger _{i\sigma\mu}  a_{j\sigma\mu} - J \sum_{i\mu} {\bf S}_{i\mu}.{\mbox{\boldmath $\sigma$}}_{i\mu}
+ U \sum_{i\mu} n_{i\mu\uparrow} n_{i\mu\downarrow} + V \sum_i n_{i\alpha} n_{i\beta} 
\end{equation}
corresponding to the two $e_g$ orbitals $\mu=\alpha,\beta$ per site, and include intra-orbital ($U$) and inter-orbital ($V$) Coulomb interactions. With increasing $V$, the orbitally degenerate ferromagnetic state becomes unstable towards an orbitally ordered ferromagnetic state with staggered orbital ordering near half-filling (in the majority-spin band). In the pseudo-spin space of the two orbitals, the orbitally-ordered state is exactly analagous to the antiferromagnetic state of the Hubbard model with staggered spin ordering.\cite{as_af_ordering} For simplicity, we consider staggered orbital ordering in all three directions, although the end compound $\rm LaMnO_3$ exhibits only planar staggered orbital ordering.

The Jahn-Teller-phononic term is also considered to be important in manganites, especially in the low and intermediate doping range.\cite{millis_1995} However, the inter-orbital Coulomb interaction has been suggested to be much stronger than the electron-phonon coupling in order to account for the observed insulating behaviour in undoped manganites above the Jahn-Teller transition and the bond length changes below it.\cite{benedetti_1999,okamoto_2002} Generally, Coulombic and Jahn-Teller-phononic approaches for manganites have been shown to be qualitatively similar.\cite{hotta_2000}
Indeed, a mean-field treatment of the Jahn-Teller term\cite{stier_2007} yields an electronic exchange-field term in orbital space proportional to the orbital magnetization $\langle n_{i\sigma\alpha} -  n_{i\sigma\beta}\rangle$, exactly as would be obtained from the inter-orbital interaction term. 

We consider an orbitally-ordered ferromagnetic state with staggered orbital ordering:
\begin{eqnarray}
\langle n_{i\alpha\uparrow}\rangle_A &=& \langle n_{i\beta \uparrow}\rangle_B = n + {\cal M}/2 \nonumber \\
\langle n_{i\beta \uparrow}\rangle_A &=& \langle n_{i\alpha\uparrow}\rangle_B = n - {\cal M}/2 
\end{eqnarray}
where the staggered orbital order ${\cal M}$ characterizes the density modulation on the two sublattices A and B. For simplicity, we consider a saturated (half-metallic) ferromagnetic state with empty spin-$\downarrow$ bands: $\langle n_{i\alpha\downarrow}\rangle = \langle n_{i\beta\downarrow}\rangle = 0$. The ferromagnetic ordering is chosen to be in the $\hat{z}$ direction.

The effective spin couplings and magnon energies are again determined from the particle-hole propagator, now evaluated in the orbitally ordered state. The magnon energies are obtained as:
\begin{equation}
\omega_{\bf q} = J^2 (2S) [\lambda(0) - \lambda({\bf q})]
\end{equation}
where $\lambda({\bf q})$ is the maximum eigenvalue of the transverse spin propagator:
\begin{equation}
\chi^{-+}({\bf q},\omega) = \sum_{\mu=\alpha,\beta} \frac{[\chi^0 _\mu ({\bf q},\omega)]}{{\bf 1} - U[\chi^0 _\mu ({\bf q},\omega)]}
\end{equation}
in terms of the bare particle-hole propagators $[\chi^0 _\mu ({\bf q},\omega)]$ for the two orbitals $\mu$, which are $2\times 2$ matrices in the two-sublattice basis, with $[\chi^0 _\beta]_{AA/BB} = [\chi^0 _\alpha]_{BB/AA}$ following from the orbital-sublattice symmetry in the orbitally-ordered state. The bare particle-hole propagators are obtained by integrating out the fermions in the orbitally ordered ferromagnetic state described by the effective single-particle Hamiltonian matrix:
\begin{equation}
H^0 _{\mu\sigma}({\bf k}) = \Gamma_\sigma \left [ \begin{array}{lr} 1 & 0 \\ 0 & 1 \end{array} \right ]
+ \left [ \begin{array}{lr} -\mu\sigma\Delta_\sigma  & \epsilon_{\bf k} \\ \epsilon_{\bf k} & \mu\sigma\Delta_\sigma \end{array} \right ]
\end{equation}
within the self-consistent field (Hartree-Fock) approximation. Here $\Gamma_\uparrow = -JS + Vn$ and $\Gamma_\downarrow = JS + (U+V)n$ are the orbitally independent fields with energy difference $\Gamma_\downarrow - \Gamma_\uparrow = Un + 2JS$ corresponding to the usual exchange band splitting, and $\Delta_\uparrow = V{\cal M}/2$ and $\Delta_\downarrow = (U-V){\cal M}/2$ are the self-consistently determined exchange fields corresponding to the orbital ordering ${\cal M}$. The spin ($\sigma$) and orbital ($\mu$) indices in the matrix are $+$ and $-$ for spins $\uparrow$ and $\downarrow$, and orbitals $\alpha$ and $\beta$, respectively. 

The eigenvalues and eigenvectors of the above Hamiltonian matrix yield the bare-level band-electron energies and amplitudes for orbital $\mu$ and spin $\sigma$:
\begin{eqnarray}
E^0 _{{\bf k}\mu\sigma} &=& \Gamma_\sigma \pm \sqrt{\Delta_\sigma ^2 + \epsilon_{\bf k}^2} \nonumber \\
a_{{\bf k}\mu\sigma \ominus}^2 &=& \frac{1}{2}\left (1 + \mu \sigma \frac{\Delta_\sigma}{\sqrt{\Delta_\sigma ^2 + \epsilon_{\bf k}^2}}\right ) = b_{{\bf k}\mu\sigma \oplus}^2 \nonumber \\
b_{{\bf k}\mu\sigma \ominus}^2 &=& \frac{1}{2}\left (1 - \mu \sigma \frac{\Delta_\sigma}{\sqrt{\Delta_\sigma ^2 + \epsilon_{\bf k}^2}}\right ) = a_{{\bf k}\mu\sigma \oplus}^2 
\end{eqnarray}
where $\oplus$ and $\ominus$ refer to the two eigenvalue branches ($\pm$). Orbital ordering splits the electron bands with energy gaps $2\Delta_\uparrow = V{\cal M}$ and $2\Delta_\downarrow = (U-V){\cal M}$ for the two spins. At quarter filling ($n=1/2$), the spin-$\uparrow$ lower band $\ominus$ is completely filled and the upper band $\oplus$ is empty, yielding a ferromagnetic insulator, and a metal-insulator transition occurs when orbital ordering melts, either driven by band overlap with decreasing $V$ or when the temperature exceeds the orbital melting temperature. In our saturated ferromagnetic state with $\langle n_{i\downarrow}\rangle = 0$, both branches of the spin-$\downarrow$ electron band are above the Fermi energy. 


The similarity of the orbitally-ordered state with the antiferromagnetic state of the Hubbard model results not only in similar expressions for quasiparticle energies and amplitudes as above, but also in an identical self-consistency condition:
\begin{equation}
\frac{1}{V} = \sum_{{\bf k}\uparrow\ominus} \frac{1}{2\sqrt{\Delta_\uparrow ^2 + \epsilon_{\bf k}^2}}
\end{equation}
which implicitly determines the magnitude of the orbital order $m=2\Delta_\uparrow/V$ as a function of the effective interaction strength $V$. When only nearest-neighbor hopping is present, orbital ordering at quarter filling sets in for any positive $V$ due to Fermi-surface nesting, whereas a finite critical interaction strength is required when frustrating next-nearest-neighbor hopping terms are included.

For finite doping $x$ away from quarter filling, the staggered orbital order parameter $m$ decreases rapidly and vanishes at a critical doping value $x_c$. Although long-range orbital order may be susceptible to orbital fluctuations at finite doping in the same way as the staggered spin ordering is in the doped antiferromagnet,\cite{as_doped_af} this HF result does provide a measure of the local orbital order. The pseudogap structure associated with short-range orbital order in the metallic phase has been observed in recent high-resolution scanning tunneling microscopy and spectroscopy measurements of 
$\rm La_{0.7} Sr_{0.3} MnO_3$ and $\rm La_{0.625} Ca_{0.375} MnO_3$ thin films.\cite{ursingh_2008} 

\begin{figure}
\begin{center}
\psfig{figure=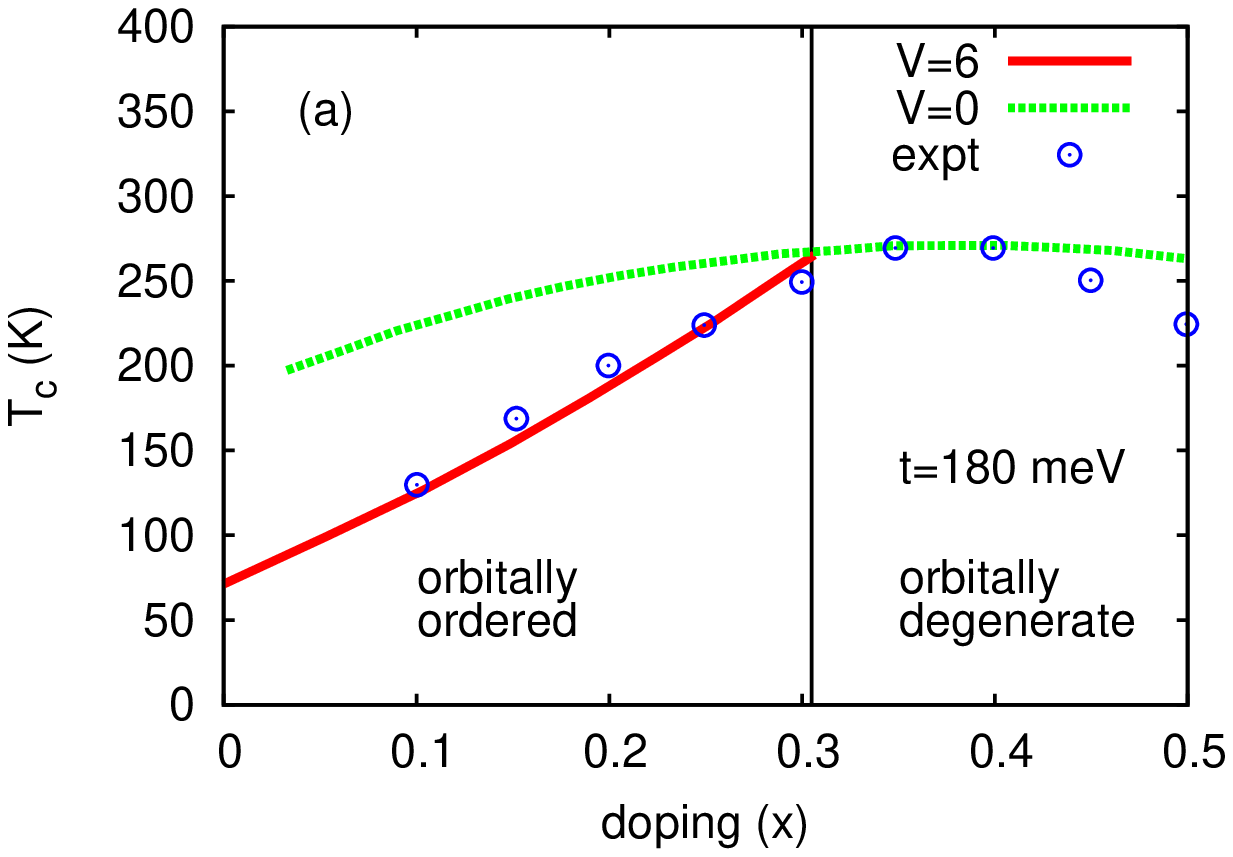,width=80mm,angle=0}
\psfig{figure=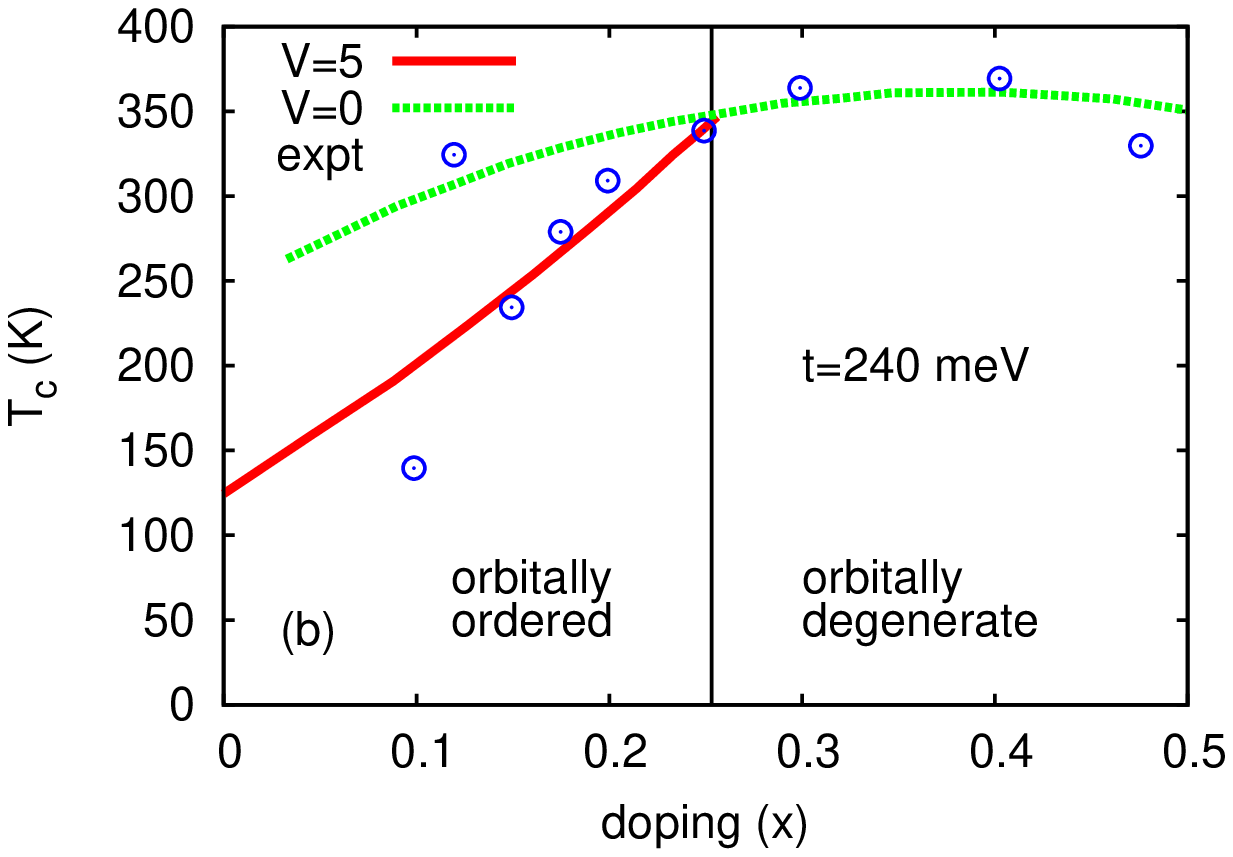,width=80mm,angle=0}
\vspace*{-5mm}
\end{center}
\caption{Comparison of the doping ($x$) dependence of calculated $T_c$ with experiments for the two compounds (a) $\rm La_{1-x} Ca_x MnO_3$ and (b) $\rm La_{1-x} Sr_x MnO_3$. Staggered orbital ordering sharply suppress ferromagnetism due to band narrowing and reduced electronic delocalization.}
\label{fig1_as}
\end{figure}

For $J\sim W$ and $U\sim W$, the Hund's coupling contribution $2JS$ to the exchange band splitting typically dominates over the intra-orbital (Hubbard) interaction contribution $Um$. Consequently, the calculated magnon energies depend rather weakly on $U$, and we have therefore dropped it for simplicity in the calculations presented below. 

Figure \ref{fig1_as} shows a comparison of the doping dependence of the calculated transition temperature $T_c$ with experiments for the two compounds $\rm La_{1-x} Ca_x MnO_3$ and $\rm La_{1-x} Sr_x MnO_3$.\cite{tc_ref1,tc_ref2} Assuming a Heisenberg form $\omega_{\bf q} = z{\cal J}S(1-\gamma_{\bf q})$ for the magnon spectrum, the transition temperature $T_c = (1/3)z{\cal J} S(S+1)/\sum_{\bf q}(1-\gamma_{\bf q})^{-1} = (5/6)\omega_{\rm X}$ was calculated approximately from the magnon energy at the X point $(\pi,0,0)$ obtained from Eq. (11). An AF superexchange interaction $J_{\rm AF}$ between the neighboring Mn core spins was also included which correspondingly reduces the magnon energy. Staggered orbital ordering was found to persist up to $x\sim 0.25$, as indicated by the dashed vertical lines, and is clearly seen to sharply suppress ferromagnetism due to band narrowing and reduced electronic delocalization. In the orbitally degenerate state obtained beyond this critical doping value, the calculated Curie temperature is indicated by $V=0$. As earlier, we have taken hopping energies $t=180$meV and $t=240$meV for the narrow-band and wide-band compounds $\rm La_{1-x} Ca_x MnO_3$ and $\rm La_{1-x} Sr_x MnO_3$, corresponding to bandwidths $\sim 2$eV and $\sim 3$eV, respectively. Other parameters are $V/t=6$ and $5$, $\rm J_{AF} = 2.8$meV and 2.1meV for the two compounds. Also, $J/t=4$ in both cases, as the bare magnon energy in the orbitally degenerate case then approximately corresponds to the renormalized magnon energy (Fig. 2) for $J\sim W$ on including the spin-charge coupling effect as discussed in section II. The anomalous zone-boundary magnon softening observed in narrow-band compounds will sharply reduce $T_c$ as $x$ approaches 0.5. 

The above analogy with the antiferromagnet can be readily extended to inter-orbital fluctuations and orbital waves (orbiton) in analogy with transverse-spin fluctuations and spin waves.\cite{as_af_ordering} This analogy then directly yields the orbiton excitation spectrum $z{\cal J}S \sqrt{1-\gamma_{\bf q}^2}$ in terms of the orbital exchange coupling ${\cal J} \equiv 4t^2/V$ in the strong-coupling limit, quantum reduction in the orbital order ${\cal M}$ due to quantum orbital fluctuations, orbital order-disorder transition at $T_c \sim (1/3)z{\cal J} S(S+1)/\sum_{\bf q} (1-\gamma_{\bf q}^2)^{-1/2}$ in the strong-coupling limit ($V\gg t$) and orbital-order melting induced metal-insulator transition in the weak-coupling limit. Similarly, hole motion in an orbitally ordered ferromagnet will lead to scrambling of the local orbital order, string of broken orbital bonds, incoherent hole spectral function and quasiparticle and band-gap renormalizations due to multiple orbiton emission-absorption processes, in analogy with corresponding results for the AF.\cite{psriv} Insights into the stability of the orbitally-ordered state at finite doping away from quarter filling can similarly be obtained from corresponding results for the doped antiferromagnet.\cite{as_doped_af} However, if realistic inter-orbital hopping terms $t_{\alpha\beta}$ are included, the continuous orbital-rotation symmetry is broken, resulting in a gapped orbiton spectrum, making the orbitally ordered state intrinsically different.

\section{CE-type orbital correlations and anomalous magnon softening}
The orbitally degenerate ferromagnetic metallic state of manganites also typically becomes unstable near hole doping $x = 0.5$ due to onset of CE-type orbital correlations. The role of such planar staggered orbital correlations with momentum near $(\pi/2,\pi/2,0)$ on spin dynamics was recently investigated for a two-orbital Hubbard model:
\begin{equation}
H = -t \sum_{\langle ij\rangle \sigma} (a^\dagger _{i\alpha\sigma} a_{j\alpha\sigma} + a^\dagger _{i\beta\sigma}  a_{j\beta\sigma} ) + U \sum_i (n_{i\alpha\uparrow} n_{i\alpha\downarrow} + n_{i\beta\uparrow} n_{i\beta\downarrow} ) 
+ \sum_i V n_{i\alpha} n_{i\beta} 
\end{equation}
including intra and inter orbital Coulomb interactions $U$ and $V$. In the saturated ferromagnetic state (band filling $n$ equal to magnetization $m$), electron correlation effects were incorporated within the framework of a non-perturbative Goldstone-mode-preserving approach.\cite{sporb} The magnon self energy due to coupling between spin and orbital fluctuations [Fig. 8] was shown to generically yield strong intrinsically non-Heisenberg $(1-\cos q)^2$ momentum dependence, implying no spin stiffness reduction but strongly suppressed zone-boundary magnon energies in the $\Gamma$-X direction. In this section we will extend these results to the two-orbital correlated FKLM as in Eq. (9) with similar manganite parameters as used in previous sections. 

The correlation-induced spin-orbital coupling magnon self energy for the two-orbital Hubbard model was obtained as:
\begin{equation}
\Sigma_{\rm sp-orb} ({\bf q}) \approx 
m^2 \frac{\langle [\Gamma_{\rm sp-orb} ({\bf q})]^2 \rangle_{{\bf Q'},\Omega'} }
{\Omega_{\rm spin} + \Omega_{\rm orb} - i \eta }
\end{equation}
where $\Omega_{\rm orb}$ and $\Omega_{\rm spin}$ represent characteristic orbital and spin fluctuation energy scales, 
the angular brackets $\langle \; \rangle$ refer to averaging over the orbital fluctuation modes ${\bf Q'}$. The spin-orbital interaction vertex $[\Gamma_{\rm sp-orb} ({\bf q})]$, represented diagrammatically in Fig. 8, was shown to explicitly vanish at momentum $q=0$ in accordance with the Goldstone mode requirement, and yields the dominant ${\bf q}$ dependence of the magnon self energy. 

\begin{figure}
\vspace*{-10mm}
\hspace*{0mm}
\psfig{figure=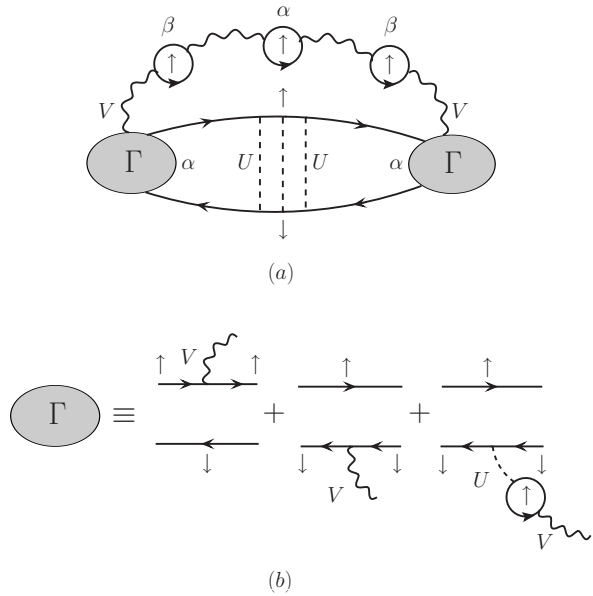,width=80mm}
\caption{The spin-orbital coupling diagrams for the irreducible particle-hole propagator (a) can be represented in terms of a spin-orbital interaction vertex $\Gamma_{\rm sp-orb}$, the three diagrammatic contributions to which are shown in (b) involving three-fermion vertices. The missing fourth diagram vanishes because of the assumption of complete polarization.}
\label{sporb}
\end{figure}

Although the same overall structure is obtained for the correlated FKLM, there are two essential differences. First, the  magnon self energy correction for the FKLM is given by $\Sigma_{\rm magnon} ^{(n)} = J^2(2S) \phi ^{(n)} $ in terms of the correponding quantum correction $\phi ^{(n)} $ to the irreducible particle-hole propagator (instead of $U^2 m \phi ^{(n)}$ as for the Hubbard model). Second, since the bare particle-hole propagator $\chi^0$ for the mobile electrons in the FKLM involves the total exchange splitting $Um + 2JS$, the spin-fluctuation propagator $\chi^0/(1-U\chi^0)$ corresponding to the $U$ ladders as in Fig. \ref{sporb} yields a gapped spectrum with characteristic gap energy $\sim 2JS$ (instead of the O($t$) magnon energy scale for the Hubbard model). 

Consequently, the first-order spin-orbital coupling magnon self energies for the two models are approximately related by:
\begin{equation}
\Sigma_{\rm sp-orb} ^{\rm FKLM} ({\bf q}) = 
\Sigma_{\rm sp-orb} ^{\rm HM} ({\bf q}) \times \left ( \frac{J^2 \; 2S}{U^2 \; m} \right ) 
\left ( \frac{t}{2JS} \right ) \; .
\end{equation} 
For $U=20t$, $m \approx 0.3$ (as taken in the Hubbard Model calculation\cite{sporb}) and $J \approx W$, the relative factor is about 1/10. For simplicity, we have therefore obtained the spin-orbital coupling magnon self energy for the FKLM from the earlier Hubbard model result by multiplying by this relative factor. 

\begin{figure}[htp]
\begin{center}
\psfig{figure=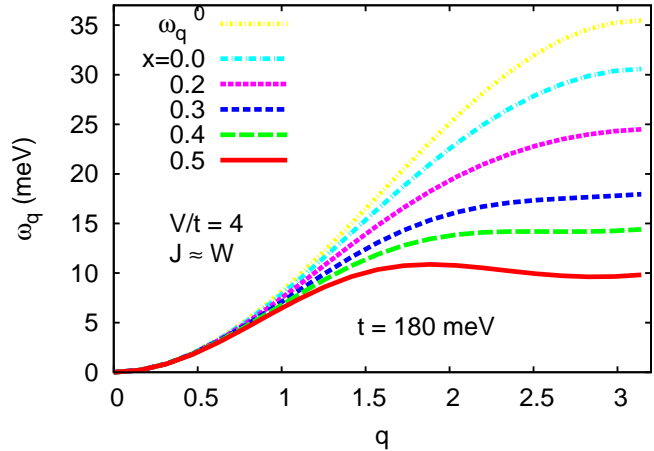,width=90mm,angle=0}
\vspace*{-5mm}
\end{center}
\caption{Renormalized magnon energies $\omega_{\bf q} = \omega_{\bf q}^0 - \Sigma_{\rm sp-orb} ({\bf q})$ including the spin-orbital coupling magnon self energy, plotted along the $\Gamma$-X direction for different hole dopings, showing the anomalous zone-boundary magnon softening due to coupling of spin excitations with CE-type, period $4a$, planar orbital correlations with momentum modes near $(\pi/2,\pi/2,0)$.}
\label{wqmev}
\end{figure}

Doubling the magnon self energy to account for the independent contributions of the two degenerate $e_g$ orbitals, the renormalized magnon energy $\omega_{\bf q} = \omega_{\bf q}^0 - \Sigma_{\rm sp-orb} ({\bf q})$ for the two-orbital correlated FKLM is shown in Fig. \ref{wqmev} for different hole dopings. Here $\omega_{\bf q}^0$ is the bare magnon energy for the two-band FKLM including the negative contribution $zJ_{\rm AF}S(1-\gamma_{\bf q})$ of the superexchange interaction $J_{\rm AF}=2.1$meV between Mn core spins. This bare magnon energy was evaluated for $J=4$ as it then approximately corresponds to the renormalized magnon energy (Fig. 2) for $J\approx W$ on including the spin-charge coupling effect as discussed in section II. 

The above analysis clearly shows the importance of CE-type staggered orbital correlations on the observed anomalous zone-boundary magnon softening. Indeed, a sharp onset of CE-type orbital correlations at momentum $(\pi/2,\pi/2,0)$ has been observed in neutron scattering studies near the Curie temperature,\cite{moussa_2007} which gradually diminishes in intensity and sharpness with decreasing Ca concentration in $\rm La_{1-x}(Ca_y Sr_{1-y})_x MnO_3$ crystals. Interestingly, this behaviour is exactly similar to the resistivity temperature profile of these crystals with varying Ca concentration.\cite{tomioka_2000} Taken in conjunction with our spin-orbital coupling theoretical result of strong zone-boundary magnon softening produced by such CE-type orbital correlations,\cite{sporb} this sharp onset of orbital correlations near $T_c$ would imply a sudden magnon softening and collapse of the ferromagnetic state. As the spin-orbital coupling magnon self energy $\sim (1-\cos q)^2$ does not renormalize the spin stiffness, this behaviour could also account for the finite spin stiffness observed just below $T_c$ in the low-bandwidth materials, which has so far been a puzzling feature. 

In the earlier Hubbard model calculation, the spin-orbital interaction term $[\Gamma_{\rm sp-orb}]^2$ was calculated for the orbital fluctuation mode ${\bf Q'}=(\pi/2,\pi/2,0)$ and an averaging factor of 1/10 was included, as obtained on averaging over all orbital fluctuation modes assuming a flat distribution. As fluctuation modes near $(\pi/2,\pi/2,0)$ become increasingly prominent due to onset of CE type combined charge-orbital correlations near half doping (stabilized by an inter-site Coulomb interaction $V' n_i n_j$), the enhanced weight of such staggered orbital correlations will substantially increase the magnon self energy and hence the anomalous magnon softening as $x$ approaches 0.5. 

\section{Conclusions}

Various aspects of spin dynamics in ferromagnetic manganites were theoretically investigated in terms of a realistic two-orbital FKLM including intra- and inter-orbital Coulomb interactions, staggered orbital ordering, and orbital correlations. For the same set of manganite parameters, the calculations are in close agreement with recent neutron scattering experiments on spin stiffness, magnon dispersion, magnon linewidth, Curie temperature, and anomalous magnon softening. The set of manganite parameters taken were lattice parameter $a=3.87$\AA, hopping terms $t=240$meV and $t=180$meV (corresponding to bandwidths $W\approx 3$eV and $\approx 2$eV, respectively) for the wide-band (LSMO) and narrow-band (LCMO) compounds, Hund's coupling $J\approx W$, inter-orbital Coulomb repulsion $V/t \approx 5$, and Mn spin superexchange interaction $J_{\rm AF}\approx 2.5$meV. 

In the orbitally degenerate ground state appropriate for the optimally doped ferromagnetic manganites with $x \sim 0.3$, the spin-charge coupling magnon self energy for the FKLM was shown to yield substantial suppression of ferromagnetism in the physically relevant finite $J$ regime. Incorporating this magnon energy renormalization, we obtained spin stiffness $D \sim 150$meV\AA$^2$, Curie temperatures $T_c \sim 300$K, and the ratio of the magnon linewidth to magnon energy $\Gamma_{\bf q}/\omega_{\bf q} \sim 1/10$ for magnon modes upto the zone boundary in the $\Gamma$-X direction, and the calculated magnon dispersion fitted well with recent neutron scattering data for $\rm La_{1-x} Ca_x MnO_3$. 

The onset of orbital ordering at low doping ($x \lesssim 0.25$) due to the inter-orbital interaction $V$ was found to sharply suppress $T_c$ due to band narrowing effects associated with the staggered orbital order. The doping dependence of the calculated Curie temperature for two bandwidth cases corresponding to narrow-band and wide-band compounds was in close agreement with the observed behaviour of $T_c$ for LCMO and LSMO.   

Finally, the role of CE-type planar orbital fluctuations with momentum near $(\pi/2,\pi/2,0)$ was investigated on spin dynamics in the orbitally degenerate ferromagnet near half doping $(x \sim 0.5)$. The magnon self energy due to correlation-induced spin-orbital coupling arising from inter-orbital interaction and fluctuation was evaluated for the correlated FKLM from the earlier Hubbard model result. For the same set of manganite parameters, the spin-orbital coupling magnon self energy with $U\sim W$ was found to yield anomalous zone-boundary magnon softening of the same magnitude as observed in neutron scattering experiments. Onset of the CE-type charge-orbital correlations and the spin-orbital coupling are therefore evidently responsible for the instability of the ferromagnetic state near half doping. 


\end{document}